\documentclass[aps,twocolumn,superscriptaddress,address,showpacs,preprintnumbers,amsmath,amssymb,prl]{revtex4}

\usepackage{graphicx}
\usepackage{dcolumn}
\usepackage{bm}

\begin{document}

\title{NMR relaxation and resistivity from rattling phonons
in pyrochlore superconductors}

\author{Thomas~Dahm}

\affiliation{Institute for Solid State Physics, University of Tokyo, 
Kashiwanoha, Kashiwa, Chiba 277-8581, Japan}

\affiliation{Institut f\"ur Theoretische Physik, 
         Universit\"at T\"ubingen, 
         Auf der Morgenstelle 14, D-72076 T\"ubingen, 
         Germany}

\author{Kazuo~Ueda}
\affiliation{Institute for Solid State Physics, University of Tokyo, 
Kashiwanoha, Kashiwa, Chiba 277-8581, Japan}

\date{\today}

\begin{abstract}
We calculate the temperature dependence of NMR relaxation rate and
electrical resistivity for coupling to a local, strongly anharmonic
phonon mode. We argue that the two-phonon Raman process is dominating
NMR relaxation. Due to the strong anharmonicity of the phonon an
unusual temperature dependence is found having a low temperature
peak and becoming constant towards higher temperatures.
The electrical resistivity is found to vary like $T^2$ at low
temperatures and following a $\sqrt{T}$ behavior at high
temperatures.
Both results are in qualitative agreement with recent
observations on $\beta$-pyrochlore oxide superconductors.
\end{abstract}


\maketitle


The recent discovery of superconductivity in the family of pyrochlore oxides 
KOs$_2$O$_6$, RbOs$_2$O$_6$, and CsOs$_2$O$_6$ has attracted great 
interest because of their unusual properties. Among these KOs$_2$O$_6$
with the highest $T_c$ of 9.6~K appears to be the most unusual.
The temperature dependence of its electrical resistivity shows a strong
concave-downward temperature dependence \cite{YonezawaK,Hiroi}, 
in contrast to the other two compounds,
where a $T^2$ temperature dependence at low temperatures has been
observed \cite{YonezawaRb,YonezawaCs}. 
Specific heat measurements have shown a large mass enhancement, 
large specific heat jump at $T_c$, 
and existence of low frequency Einstein modes
\cite{Bruhwiler,Hiroirev}.
Bandstructure calculations have indicated that the anomalies, in
particular in KOs$_2$O$_6$, might be due to a highly anharmonic
low frequency rattling motion of the alkali-ion inside an oversized cage
formed by the Os and O ions \cite{Kunes1,Kunes2}. 
This is consistent with
X-ray observations of anomalously large atomic displacements for the K 
ions \cite{Yamaura} and low frequency phonon structures seen
in photoemission spectroscopy \cite{Shin}.

Recent observations of NMR relaxation rates $1/T_1T$
at the K site have been
demonstrated to be entirely dominated by the vibrations of the K ion
via coupling of the electric field gradient to the nuclear
quadrupole moment \cite{Yoshida}. Such a domination of phonons for nuclear
relaxation usually occurs in diamagnetic insulators, but is
extremely rare in metals and was attributed to the rattling motion
of the K ions. The temperature dependence of $1/T_1T$
was found to be anomalous as well, showing a peak around 12-14 K
and decreasing at higher temperatures. It has been argued that
such a behavior is inconsistent with the two-phonon Raman process,
which usually dominates quadrupolar relaxation, and it has been
interpreted in terms of the direct phonon process with a strongly
temperature dependent phonon damping rate \cite{Yoshida}. 
In the superconducting state $1/T_1T$ exhibits a sudden decrease,
suggesting a strong coupling of the phonon mode to the conduction
electrons and an associated increase of the phonon lifetime in the
superconducting state.

Motivated by these experimental findings, in the present work we
study the influence of a local, strongly anharmonic, and
damped phonon mode on the NMR relaxation rate and the electrical
resistivity. We find that within this model the two-phonon Raman
process is expected to dominate NMR relaxation over the direct process.
Due to the anharmonicity of the phonon mode the
temperature dependence of $1/T_1T$ for the Raman process is
qualitatively different from harmonic phonons.
We show that the experimental data are qualitatively reproduced,
showing a peak at about half the low temperature phonon frequency
and decreasing towards a constant value at higher temperatures.
Within the same model also the temperature dependence of the
resistivity can be understood, following a $T^2$ law at low
temperatures and a $\sqrt{T}$ behavior at high temperature.
Thus, a qualitative understanding of both the anomalous NMR relaxation
and electrical resistivity can be obtained simultaneously by considering the
anharmonicity of the phonon involved.

As a description of the local alkali-ion motion we start from an 
anharmonic Hamiltonian of the form
\begin{equation}
H = \frac{p^2}{2 M} + \frac{1}{2} a x^2 + \frac{1}{4} b x^4 
\label{hamil1}
\end{equation}
where $x$, $p$, and $M$ are the spacial coordinate, momentum, and
mass of the alkali-ion, respectively, and $a$ and $b>0$ are constants.
We note that according to bandstructure calculations, in KOs$_2$O$_6$ 
the quadratic term even becomes negative $a<0$, resulting in a shallow 
double well potential, while in the other two compounds $a$ is 
positive \cite{Kunes1}. We treat this Hamiltonian
in a selfconsistent quasi-harmonic approximation resulting in
an effective harmonic Hamiltonian
\begin{equation}
H = \frac{p^2}{2 M} + \frac{1}{2} M \omega_0^2 x^2
\label{hamileff}
\end{equation}
where the effective phonon frequency $\omega_0$ becomes
temperature dependent and has to be determined selfconsistently
from the equation
$M \omega_0^2 = a + b \langle x^2 \rangle_{\omega_0,T}$.
Here, $\langle x^2 \rangle_{\omega_0,T}$ denotes the thermal
average of $x^2$ that depends on both $\omega_0$ and temperature $T$
and is given by
\begin{equation}
\langle x^2 \rangle_{\omega_0,T} = \frac{\hbar}{M \omega_0}
\left[ \frac{1}{e^{\hbar \omega_0 / k_B T}-1} + \frac{1}{2} \right] \; .
\end{equation}
\begin{figure}[t]
\includegraphics[width=0.75 \columnwidth, angle=270]{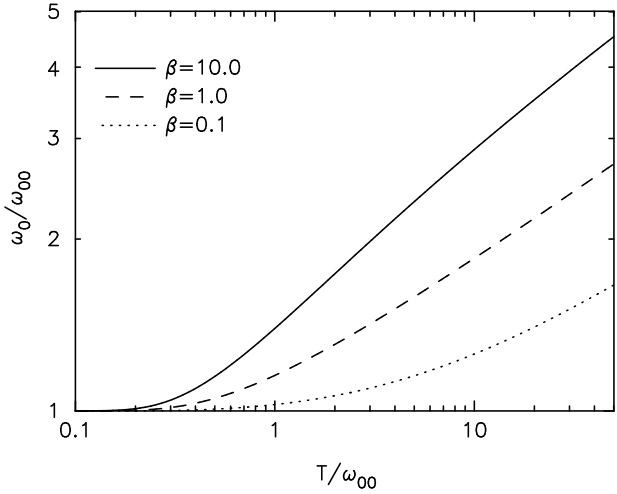}
\caption{\label{fig1}
Effective phonon frequency $\omega_0/\omega_{00}$ as a function
of temperature $T / \omega_{00}$ for 
anharmonicity parameter $\beta=$~0.1, 1, and 10 (double logarithmic scale).
}
\end{figure}
Note, that the effective phonon frequency $\omega_0$ is a thermodynamical
average frequency and not to be confused with the discrete, temperature
independent energy levels of the Hamiltonian Eq.~(\ref{hamil1}).
In the zero temperature limit we have
\begin{equation}
\omega_0^2(T=0) = \frac{a}{M} + b \frac{\hbar}{2 M^2 \omega_0(T=0)}
\label{omega00}
\end{equation}
Note that this equation guarantees $\omega_0(T=0)>0$ for $b>0$,
even if $a<0$. In the high temperature limit $T \gg \omega_0$
we find
\begin{equation}
\omega_0 \sim \left( \frac{b k_B T}{M^2} \right)^{1/4} \; .
\label{tquarter}
\end{equation}
We can eliminate the parameter $a$ in favor
of $\omega_{00}=\omega_0(T=0)$ using Eq.~(\ref{omega00}). Then
we can derive the following nonlinear equation:
\begin{equation}
\left( \frac{\omega_0}{\omega_{00}} \right)^2 = 1 + \beta \frac{\omega_{00}}{\omega_0}
\left[ \frac{1}{e^{\hbar \omega_0 / k_B T}-1} + \frac{1}{2} - \frac{1}{2} \frac{\omega_0}{\omega_{00}} \right]
\label{efffreq}
\end{equation}
where $\beta = b \frac{\hbar}{M^2 \omega_{00}^3} > 0$ is a dimensionless parameter
characterizing the amount of anharmonicity. In Fig.~\ref{fig1}
$\frac{\omega_0}{\omega_{00}}$ is shown as a function of
$T / \omega_{00}$ for $\beta=$~0.1, 1, and 10.
$\frac{\omega_0}{\omega_{00}}$ is a monotonously increasing 
function of temperature.

\begin{figure}[t]
  \begin{center}
    \includegraphics[width=0.9\columnwidth]{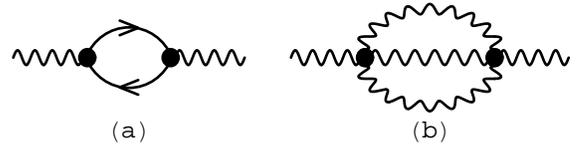}
    \caption{Two lowest order processes leading to a finite phonon
lifetime: (a) coupling of the phonon to the conduction electrons and
(b) fourth order phonon process due to the anharmonicity. \label{fig2} }
  \end{center}
\end{figure}

Defining the retarded phonon propagator as
\begin{equation}
D(\omega) = - i \frac{2 \omega_0 M}{\hbar} \int_0^\infty dt \; e^{i \omega t} 
\left\langle \left[ x \left( t \right), 
x \left( 0 \right) \right] \right\rangle
\label{phononprop}
\end{equation}
the interaction of the phonon with the conduction electrons can
be described by the phonon self-energy $\Pi(\omega)$ via
Dyson's equation
$D^{-1}(\omega) = D_0^{-1}(\omega)- \Pi(\omega)$.
Here, the noninteracting phonon propagator due to the
effective Hamiltonian Eq.~(\ref{hamileff}) is given by
$
D_0(\omega) = \frac{2 \omega_0 }{\omega^2 - \omega_0^2} 
$
Thus, the phonon propagator can be written in terms of
the self-energy as
\begin{equation}
D(\omega) = \frac{2 \omega_0 }{\omega^2 - \omega_0^2 - 2 \omega_0 \Pi(\omega)} 
\end{equation}
The real part of the self-energy leads to a renormalization of
the phonon frequency
$\omega_r^2 =  \omega_0^2 + 2 \omega_0 \mathrm{Re} \; \Pi(\omega)$
while the imaginary part determines the damping width
$\Gamma (\omega)  = - \mathrm{Im} \; \Pi(\omega)$.
Due to symmetry the damping width has to be an odd
function $\Gamma (\omega)  = - \Gamma (- \omega)$, while 
the real part has to be even
$\mathrm{Re} \; \Pi(\omega) = \mathrm{Re} \; \Pi(- \omega)$.
In Fig.~\ref{fig2} we show the two lowest order processes
contributing to a finite phonon damping. Process (b) is due to
the fourth order process neglected in Eq.~(\ref{hamileff}).
This process does not lead to a change of the phonon
damping, when superconductivity sets in. As mentioned above,
the NMR results suggest that there is a strong coupling of the
phonon to the conduction electrons and therefore we expect process (a)
to play the leading role.
Since the Fermi energy of the conduction electrons is a much
larger energy scale than the phonon energies, process (a) will
be featureless in the low energy range of interest in the normal
state. The damping width is a linear 
function at low energy and we can write
$\Gamma (\omega)  = \alpha \omega$.
The real part of the self-energy can be assumed constant
and must be negative (i.e. the electron-phonon interaction
leads to a softening of the phonon).
The dimensionless parameter $\alpha$ determines the phonon
damping and is given by $\alpha = g^2 N^2(0)$, where
$N(0)$ is the density of states at the
Fermi level and $g$ the electron-phonon coupling constant.
It is important to keep in mind that $g^2 \sim \omega_0^{-1}$,
because here $\omega_0$ is temperature dependent. Therefore,
in the following we write $\alpha=\Gamma_0/\omega_0$,
where $\Gamma_0$ is a (temperature independent) phonon damping 
rate and will be treated as a phenomenological parameter.
The phonon spectral function is given by
\begin{equation}
A (\omega) = - \frac{1}{\pi}  \mathrm{Im} \; D(\omega)
= \frac{1}{\pi} \frac{4 \omega_0 \Gamma_0 \omega}{
\left( \omega^2 - \omega_r^2 \right)^2 + 4 \Gamma_0^2 \omega^2}
\label{phnspcfct}
\end{equation}
This is equivalent to the familiar damped harmonic oscillator
formula apart from the fact that $\omega_r$ becomes temperature
dependent here due to the anharmonicity.

The quadrupolar NMR relaxation rate is usually dominated by the
two phonon Raman process \cite{Abragam}, which is given by
\begin{equation}
\frac{1}{T_1^R} = V_2^2
\int_{-\infty}^\infty dt \; e^{i \omega_L t} 
\left\langle x^2 \left( t \right)
x^2 \left( 0 \right) \right\rangle
\label{Ramanprocess}
\end{equation}
where $V_2$ is proportional to the second spacial derivative of
the electric field gradient and $\omega_L$ is the nuclear Lamor frequency,
which is usually very small. In Ref. \onlinecite{Yoshida} it has
been suggested, however, that in the case of KOs$_2$O$_6$
the direct (one phonon) process might be dominating instead.
Here, we are comparing these two processes for our model.
For the direct process we have
\begin{equation}
\frac{1}{T_1^D} = V_1^2
\int_{-\infty}^\infty dt \; e^{i \omega_L t} 
\left\langle x \left( t \right)
x \left( 0 \right) \right\rangle
\end{equation}
where $V_1$ is proportional to the first spacial derivative of
the electric field gradient. Using the fluctuation-dissipation
theorem we can directly relate $\frac{1}{T_1^D}$ to the
phonon spectral function:
$
\frac{1}{T_1^D} = 2 \pi V_1^2
\frac{\hbar}{2 \omega_0 M} 
\frac{1}{ 1 - e^{-\hbar \omega_L / k_B T} } A (\omega_L)
$.
Assuming $\hbar \omega_L \ll k_B T$ we find
\begin{equation}
\frac{1}{T_1^D T} = 2 \pi V_1^2
\frac{k_B}{2 \omega_0 M} 
\lim_{\omega \rightarrow 0} \frac{A(\omega)}{\omega} 
= 2 V_1^2
\frac{k_B}{2 \omega_0 M} 
\frac{4 \omega_0 \Gamma_0}{\omega_r^4 }
\label{directhigh}
\end{equation}
As has been pointed out in Ref.~\onlinecite{Yoshida},
this expression is monotonously decreasing as a
function of temperature and thus cannot reproduce the
observed peak in $\frac{1}{T_1 T}$.
Due to the temperature dependence of $\omega_0$ at high
temperatures we have
$ \frac{1}{T_1^D T} \sim \omega_0^{-4} \sim T^{-1}$.

Following the discussion of the Raman process in 
Ref.~\onlinecite{Abragam} we can also express 
Eq.~(\ref{Ramanprocess}) in terms of the phonon
spectral function:
\begin{equation}
\frac{1}{T_1^R} = 2 \pi \left( \frac{\hbar}{2 \omega_0 M} 
\right)^2 V_2^2
\int_{-\infty}^\infty d\omega \; A^2 (\omega)
\left[ n( \omega ) + 1 \right] n (\omega )
\label{Ramanprocess2}
\end{equation}
At sufficiently high temperatures $T \gg \omega_r$, in the
normal state we can approximate the Bose function
$n (\omega ) \approx \frac{k_B T}{\hbar \omega}$
and Eq.~(\ref{Ramanprocess2}) can be evaluated analytically.
After some algebra we arrive at
\begin{equation}
\frac{1}{T_1^R} =
\left( \frac{k_B T V_2}{2 \omega_0 M} 
\right)^2 \frac{2\omega_0^2}{\Gamma_0}
\frac{4\Gamma_0^2 +\omega_r^2}{\omega_r^6}
\label{Ramanhigh}
\end{equation}
Due to the temperature dependence of $\omega_0$ in Eq.~(\ref{tquarter})
we find the high temperature behavior
\begin{equation}
\frac{1}{T_1^R T} \sim T \omega_0^{-4} \sim {\mathrm{const}}
\label{Ramanasym}
\end{equation}
Thus, $\frac{1}{T_1^R T}$ approaches a constant high
temperature value for anharmonic phonons. This is in
strong contrast to harmonic phonons, where $\frac{1}{T_1^R T}$ is
increasing linearly with $T$.
To estimate the relative importance of the two processes
we can use Eqs.~(\ref{Ramanhigh}) and (\ref{directhigh})
to calculate the ratio
\begin{equation}
\frac{T_1^D}{T_1^R} =
\frac{V_2^2}{V_1^2} \frac{k_B}{2 \omega_0 M}  
\frac{4\Gamma_0^2 +\omega_r^2}{\omega_r^2}
\frac{T \omega_0}{4 \Gamma_0^2}
\end{equation}
From this expression it becomes clear that the Raman
process is certainly dominating at high temperature and
also, when the phonon damping $\Gamma_0$ is small.
We note that the ratio $\frac{V_2}{V_1}$ has dimension
of an inverse length, which should be of the order of
magnitude of the size of the potential, in which the
K ion moves. Therefore, as an order of magnitude estimate
we assume
$\frac{V_2^2}{V_1^2} \frac{\hbar}{2 \omega_{00} M} \sim 1  $

\begin{figure}[t]
\includegraphics[width=0.75 \columnwidth, angle=270]{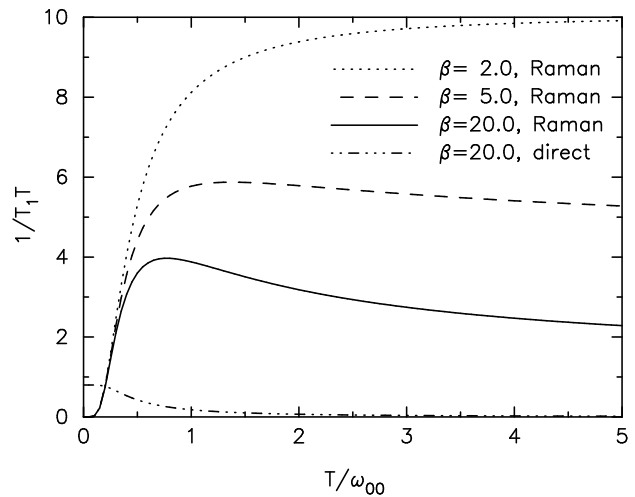}
\caption{\label{fig3}
Temperature dependence of $1/T_1T$ for the Raman process and
anharmonicity parameter $\beta=$~2 (dotted), 5 (dashed), and 20
(solid). The dashed-dotted line shows the direct process for
comparison.
}
\end{figure}

In the normal state at low enough temperatures Eq.~(\ref{Ramanprocess2})
leads to
\begin{equation}
\frac{1}{T_1^R T} \sim T^2
\end{equation}
because of the linear $\omega$ dependence of $A(\omega)$.
As an illustration Fig.~\ref{fig3} shows
a numerical calculation of $\frac{1}{T_1^R T}$ as a function
of $T/\omega_{00}$ based on Eqs.~(\ref{Ramanprocess2}) and
(\ref{efffreq}) for $\Gamma_0/\omega_{00}=0.1$, $\mathrm{Re} \; \Pi =0$, and
three different values for the anharmonicity parameter $\beta=$~2, 5, and 20.
Clearly, with increasing anharmonicity a low temperature peak
develops near $T \approx \omega_{00}/2$ and becomes more pronounced. 
This can be understood
from the $\omega_0^{-4}$ behavior in Eq.~(\ref{Ramanasym}):
at low temperatures $\omega_0$ becomes constant and all curves
fall on top of each other. However, at higher temperatures
$\omega_0$ increases the stronger the larger $\beta$ is,
resulting in a quick decrease for larger $\beta$.
For comparison, the dashed-dotted line shows the direct process
for $\beta=20$, assuming 
$\frac{V_2^2}{V_1^2} \frac{\hbar}{2 \omega_{00} M}=1$.
Clearly, the direct process can be neglected
over most of the temperature range. We mention that the 
direct process is exponentially
suppressed in the superconducting state.

\begin{figure}[t]
\includegraphics[width=0.75 \columnwidth, angle=270]{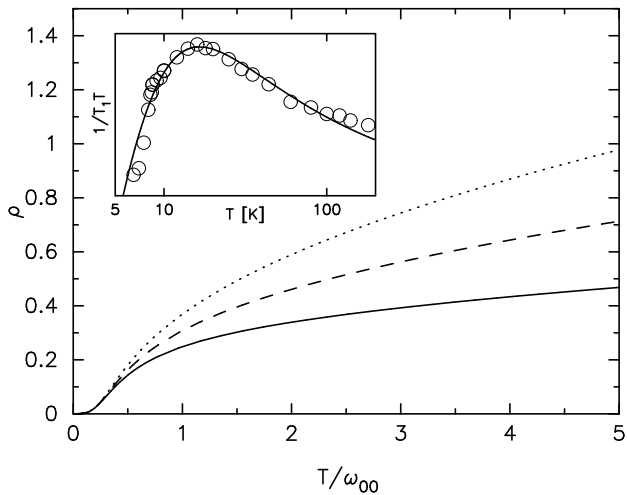}
\caption{\label{fig4}
Temperature dependence of the resistivity for the same parameters
as in Fig.~\ref{fig3}. Inset: 
Temperature dependence of $1/T_1T$ fitted to the experimental
data of Ref.~\onlinecite{Yoshida} (sample A) on a double
logarithmic scale.
}
\end{figure}

In the inset of Fig.~\ref{fig4} we show a fitting of Eq.~(\ref{Ramanprocess2})
to the experimental data of Ref.~\onlinecite{Yoshida} (sample A).
From this fit the following parameters are found:
$\omega_r(T=0)=27.4$~K, $\beta=6.27$, $\Gamma_0=4.0$~K, and
$\omega_0 \mathrm{Re} \; \Pi =$-(21~K)$^2$. The low temperature value of the
renormalized phonon frequency $\omega_r(T=0)=27.4$~K is in reasonable
agreement with a low frequency Einstein frequency of 22~K found in
fits to specific heat data \cite{Hiroirev} and 24.4~K found from
a phonon structure in photoemission \cite{Shin}.

In the following we calculate the resistivity due
to coupling to the local anharmonic phonon Eq.~(\ref{phnspcfct}).
We will follow the work by Mahan and Sofo \cite{Mahan}.
The imaginary part of the retarded electronic self energy is
given by
\begin{eqnarray}
\lefteqn{- \mathrm{Im} \; \Sigma(E) = \tau^{-1}(E) = } 
\label{elselfen} \\ & &
\pi g^2 N(0) \int_0^\infty d\Omega \;
A(\Omega) \left[ 2 n \left( \Omega \right) 
+ f \left( \Omega + E \right) + f \left( \Omega - E \right)\right]
\nonumber
\end{eqnarray}
where $f(\Omega)=\frac{1}{e^{\hbar \omega / k_B T}+1}$ is the Fermi 
function. The resistivity can be calculated from 
$\rho(T) = \frac{4\pi}{\omega_p^2} \frac{1}{\tau(T)}$
where $\omega_p$ is the Plasma frequency and the electronic 
lifetime is given by
$\tau(T) = - \int_{-\infty}^\infty dE \; \tau(E) \frac{df}{dE}$.
We note that we can neglect vertex corrections here, because
the phonon is local, i.e. momentum independent. Therefore,
the transport lifetime is the same as the electronic lifetime.

Because $A(\Omega)$ is linear in $\Omega$ at low frequencies,
we find from Eq.~(\ref{elselfen}) that at temperatures
$T \ll \omega_r$ the resistivity varies like $\rho \sim  T^2$.
However, in the high temperature limit we find from Eq.~(\ref{elselfen}):
\begin{equation}
\tau^{-1} = 
2 \pi T g^2 N(0) \int_0^\infty d\Omega \;
\frac{A(\Omega)}{\Omega}= 
2 \pi T \frac{\Gamma_0}{N(0) \omega_r^2}
\end{equation}
Therefore, at high temperatures the resistivity
varies as $\rho \sim \sqrt{T}$, a concave-downward
temperature dependence.
In Fig.~\ref{fig4} numerical results for the resistivity $\rho$
are shown for the same parameters as in Fig.~\ref{fig3}. With
increasing $\beta$ the concave-downward
temperature dependence becomes more pronounced.
Considering $\omega_{00} \sim 25$~K this result is in qualitative
agreement with the resistivity data on KOs$_2$O$_6$ 
\cite{Hiroi,Hiroirev}. A quantitative fitting
would require taking into account the other, harmonic phonon modes
as well. Since at present not much is known about these a more
quantitative comparison has to await more detailed experimental
information.

In conclusion, we have calculated the temperature dependence of
NMR relaxation rates and resistivity due to coupling to a strongly
anharmonic phonon mode. Our results are in agreement with recent
experiments on $\beta$-pyrochlore superconductors, where such an
anharmonic rattling motion of the alkali-ions is believed to
cause various anomalies. Our theory may also be applicable to other
rattling systems like the filled skutterudite LaOs$_4$Sb$_{12}$, as a
recent NMR study suggests \cite{Nakai}.

Thanks are due to M.~Takigawa and Z.~Hiroi for valuable
discussions and for providing their data.

\end{document}